\newcommand{\kaonangle}{$\cos\theta_\mathrm{CM}^{K}$}
\documentclass[final,5p,times]{elsarticle}

\usepackage{amssymb}
\usepackage{graphics}
\usepackage{graphicx}
\usepackage{xcolor}
\usepackage{sidecap}
\usepackage{microtype}
\usepackage{siunitx}

\usepackage{lineno}

\journal{Physics Letters B}

\begin{document}

\begin{frontmatter}



\title{Observation of a cusp-like structure in the $\gamma p \rightarrow K^+\Sigma^0$ cross section at forward angles and low momentum transfer}

\author[1]{T.C.~Jude\corref{cor1}}
\ead{jude@physik.uni-bonn.de}
\author[1]{S.~Alef}
\author[1]{P.~Bauer}
\author[2,3]{D.~Bayadilov}
\author[2]{R.~Beck}
\author[1]{A.~Bella\fnref{fn1}}
\author[2]{J.~Bieling\fnref{fn1} }
\author[4]{A.~Braghieri}
\author[5]{P.L.~Cole}
\author[1]{D.~Elsner}
\author[6]{R.~Di Salvo}
\author[6,7]{A.~Fantini}
\author[1]{O.~Freyermuth}
\author[1]{F.~Frommberger}
\author[8,9]{F.~Ghio}
\author[1]{S.~Goertz}
\author[3]{A.~Gridnev}
\author[1]{D.~Hammann\fnref{fn1} }%
\author[1]{J.~Hannappel \fnref{fn2} }
\author[1]{K.~Kohl}
\author[3]{N.~Kozlenko}
\author[10]{A.~Lapik}
\author[11]{P.~Levi Sandri}
\author[10]{V.~Lisin}
\author[12,13]{G.~Mandaglio}
\author[1]{F.~Messi \fnref{fn1} }%
\author[6,7]{R.~Messi}
\author[11]{D.~Moricciani}
\author[10]{V.~Nedorezov}
\author[2,3]{V.A~Nikonov  \fnref{fn3} }
\author[3]{D.~Novinskiy}
\author[4]{P.~Pedroni}
\author[10]{A.~Polonskiy}
\author[1]{B.-E.~Reitz \fnref{fn1} }%
\author[6,14]{M.~Romaniuk}
\author[2,3]{A.V~Sarantsev}
\author[1]{G.~Scheluchin}
\author[1]{H.~Schmieden}
\author[3]{A.~Stuglev}
\author[3]{V.~Sumachev  \fnref{fn3} }
\author[1]{V.~Vegna \fnref{fn1} }%
\author[3]{V.~Tarakanov}
\author[1]{T.~Zimmermann\fnref{fn1} }

\cortext[cor1]{Corresponding author}
\fntext[fn1]{No longer employed in academia}
\fntext[fn2]{Currently, DESY Research Centre, Hamburg, Germany}
\fntext[fn3]{Deceased}

\address[1]{Rheinische Friedrich-Wilhelms-Universit\"at Bonn, Physikalisches Institut, Nu\ss allee 12, 53115 Bonn, Germany}
\address[2]{Rheinische Friedrich-Wilhelms-Universit\"at Bonn, Helmholtz-Institut f\"ur Strahlen- und Kernphysik, Nu\ss allee 14-16, 53115 Bonn, Germany}
\address[3]{Petersburg Nuclear Physics Institute, Gatchina, Leningrad District, 188300, Russia}
\address[4]{INFN sezione di Pavia, Via Agostino Bassi, 6 - 27100 Pavia, Italy}
\address[5]{Lamar University, Department of Physics, Beaumont, Texas, 77710, USA}
\address[6]{INFN Roma ``Tor Vergata", Via della Ricerca Scientifica 1, 00133, Rome, Italy}
\address[7]{Universit\`a di Roma ``Tor Vergata'', Dipartimento di Fisica, Via della Ricerca Scientifica 1, 00133, Rome, Italy}
\address[8]{INFN sezione di Roma La Sapienza, P.le Aldo Moro 2, 00185, Rome, Italy}
\address[9]{Istituto Superiore di Sanit\`a, Viale Regina Elena 299, 00161, Rome, Italy}
\address[10]{Russian Academy of Sciences Institute for Nuclear Research, Prospekt 60-letiya Oktyabrya 7a, 117312, Moscow, Russia}
\address[11]{INFN - Laboratori Nazionali di Frascati, Via E. Fermi 54, 00044, Frascati, Italy}
\address[12]{INFN sezione Catania, 95129, Catania, Italy}
\address[13]{Universit\`a degli Studi di Messina, Dipartimento MIFT,  Via F. S. D'Alcontres 31, 98166, Messina, Italy}
\address[14]{Institute for Nuclear Research of NASU, 03028, Kyiv, Ukraine}

\begin{abstract}
		The $\gamma p \rightarrow K^+\Sigma^0$  differential cross section at extremely forward angles was measured at the BGOOD experiment. A three-quarter drop in strength over a narrow range in energy and a strong dependence on the polar angle of the $K^+$ in the centre-of-mass of the reaction is observed at a centre-of-mass energy of 1900\,MeV. Residing close to multiple open and hidden strangeness thresholds, the structure appears consistent with meson-baryon threshold effects which may contribute to the reaction mechanism.
\end{abstract}




\end{frontmatter}


\section{Introduction}

Recent discoveries of ``exotic" multi-quark hadronic states beyond the conventional three and two constituent quark models have challenged the understanding of the degrees of freedom afforded in QCD.
 In the charmed sector this includes for example, the $P_c$ pentaquark states~\cite{aaij15,aaij19}
and a plethora of XYZ mesons~\cite{chen16}, and in the light quark sector, the $\Lambda$(1405)~\cite{guo18} and $d^*(2380)$ hexaquark~\cite{bashkanov09}.

Descriptions of multi-quark states have existed since the conception of quark models~\cite{gellmann64,jaffe77,strottmann79}, and    
due to the proximity of the chiral symmetry breaking scale to the nucleon mass, 
it is possible that light mesons effectively interact as elementary objects, giving rise to molecular systems and meson re-scattering effects near thresholds~\cite{manohar84,glozman96}.
Models including meson-baryon interactions and dynamically generated states~\cite{dalitz67,siegel88,kaiser97,recio04,lutz04,nacher03} have had improved success in describing nucleon excitation spectra.
In the strangeness sector, the $\Lambda$(1405) is considered a molecular $\bar{K}N$ state to some extent~\cite{nacher03,dalitz59}, which is also supported by Lattice QCD calculations~\cite{hall15}. 
Similar states may have been observed in $K^0\Sigma^+$ photoproduction, where a ``cusp" was observed where the cross section quickly reduces by a factor of 75\,\% at the $K^*\Lambda$ and $K^*\Sigma^0$ thresholds~\cite{ewald12,ewald14}. 
This was described by the destructive interference of amplitudes driven by intermediate $K^*\Lambda$ and $K^*\Sigma$ channels~\cite{oset13}.  
Other hadronic molecules with hidden and open strangeness have been proposed.  This includes molecular $K\Sigma$ states with masses and quantum numbers ($IJ^P = \frac{1}{2}\frac{1}{2}^-$ and $\frac{1}{2}\frac{3}{2}^-$)  consistent with the $N^*(1875)$ and $N^*(2100)$ resonances~\cite{huang18}, bound  $\phi N$ systems with a mass of 1950\,MeV just below the $p\phi$ threshold~\cite{guo17} and a three-hadron $K\bar{K}N$ molecule comprised of $a_0(980)N$ and $f_0(980)N$ components with $J^P = \frac{1}{2}^+$~\cite{torres09}.  It is interesting to note the almost mass degeneracy of these predictions close to 1900\,MeV and close to numerous thresholds, including $K^+K^-p$, $K\Sigma(1385)$, $K^+\Lambda$(1405), $f_0$(980)$p$, $a_0$(980)$p$  and $\phi p$.

To experimentally verify the formation of any such extended, loosely bound molecular systems requires minimal momentum transfer kinematics.
  For fixed target photoproduction experiments,  access to forward meson angles and corresponding low momentum transfer to the recoiling baryon is therefore crucial.  
  
  This letter presents differential cross section data for $\gamma p \rightarrow K^+\Sigma^0$ from threshold at forward angles, where the cosine of the centre-of-mass polar angle of the $K^+$ (\kaonangle{}) is between 0.9 and 1.0.
  There is limited data available for this reaction in this kinematic regime.  The CLAS datasets only extend to \kaonangle{}$<0.95$~\cite{dey10,bradford06}, and do not agree in this most forward angle interval.  The SAPHIR data~\cite{glander04} extend to \kaonangle{} $ = 1.0$ however is significantly lower than the CLAS data and with lower statistical precision.
The only data with $K^+\Sigma^0$ in the final state in this kinematic regime are from the Sphinx Collaboration, where they reported preliminary strange pentaquark evidence in the diffractive production: $ p + C \rightarrow [Y^0K^+] + C$~\cite{golovkin95}.  The proposed $X(2000)$ was observed in the $K^+\Sigma^0$ invariant mass spectra with a width of 91\,MeV.  To ensure a coherent process, the Sphinx Collaboration required a transverse momentum of the $K^+\Sigma^0$ system to be lower than $300$\,MeV/c.
Crucially in photoproduction experiments, access to small transverse momenta when away from threshold also requires very forward meson acceptance.  This has not been achieved with any existing photoproduction measurements and indeed a review in 1997 could not confirm the $X(2000)$ has been observed in the CLAS data~\cite{schumacher97} .

The presented $\gamma p \rightarrow K^+\Sigma^0$ cross section was measured at the BGOOD experiment~\cite{technicalpaper} at the ELSA electron accelerator facility~\cite{hillert06,hillert17}.  This is the first dataset with the statistical precision, forward $K^+$ acceptance and resolution to describe the data trend in this low momentum transfer region.
A cusp-like drop in strength is observed at extremely forward angles at a centre-of-mass energy of 1900\,MeV.

%

\section{Experimental setup and analysis procedure}

BGOOD is composed of a \textit{Forward Spectrometer} for charged particle identification and momentum reconstruction over laboratory frame polar angles $1^\circ$ to $12^\circ$.
This is complemented by the \textit{BGO Rugby Ball}, an almost $4\pi$ central calorimeter ideal for photon detection with sub-detectors for charged particle identification.
The experimental conditions and $K^+$ identification are described in refs.~\cite{technicalpaper,klambdapaper}.

\begin{figure} [h]
	\centering
	\includegraphics[width=1.0\columnwidth]{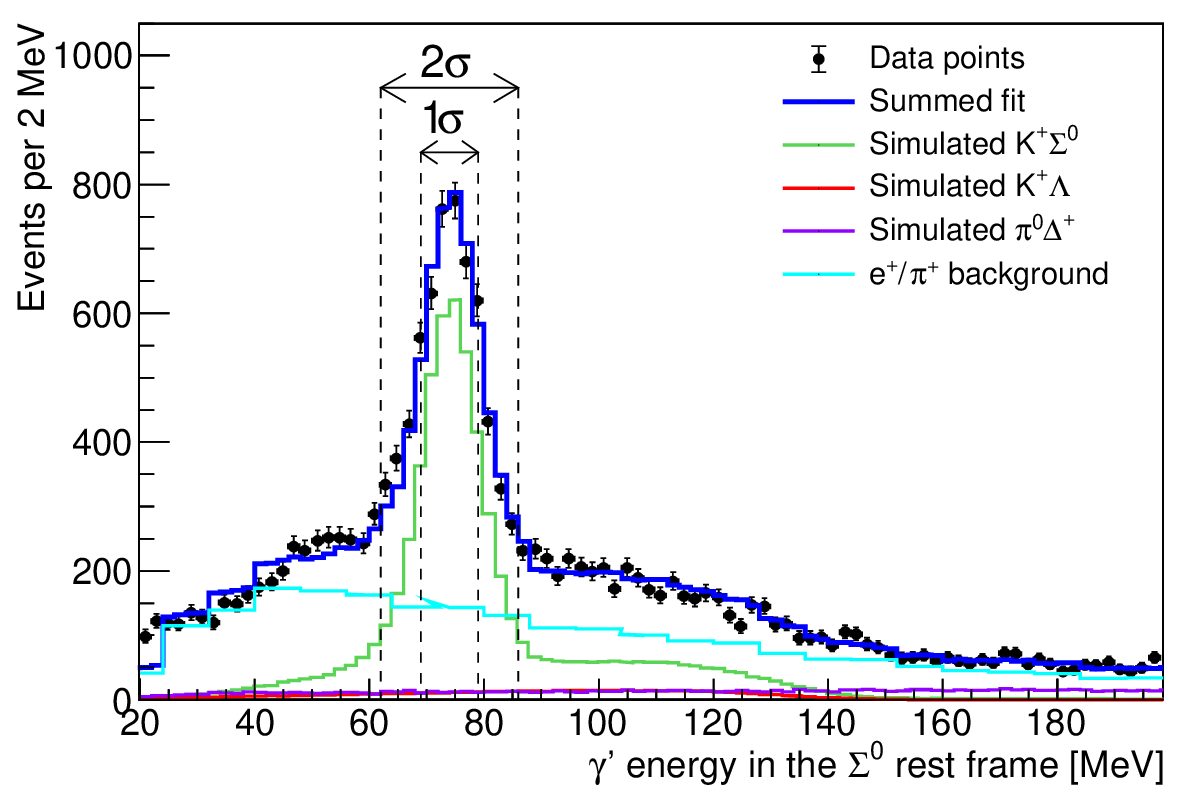} 	
	\caption{$\gamma'$ energy in the $\Sigma^0$ rest frame, with a peak consistent with the expected energy from the $\Sigma^0\rightarrow \Lambda\gamma$ decay.  Simulated spectra of $K^+\Sigma^0$ (green line), $K^+\Lambda$ (red line), $\Delta^0\pi^+$ (magenta line)  and $e^+$/$\pi^+$ background (cyan line) are fitted to the data.  The summed total fit is the blue line.  The dashed vertical lines show the $1\sigma$ and $2\sigma$ selection cuts described in the text.}
	\label{fig:decayphoton}
\end{figure}

The data were taken over 22 days using an electron beam energy of 3.2\,GeV and a 6\,cm long liquid hydrogen target.   The electron beam was incident upon a thin diamond radiator to produce an energy tagged bremsstrahlung photon beam which was subsequently collimated\footnote{A diamond radiator was used to produce coherent, linearly polarised photon beam with a maximum polarisation at a beam energy of 1.4\,GeV, however the polarisation was not required and was averaged out for the presented analysis.}.  The photon beam energy, $E_\gamma$, was determined per event by momentum analysing the post bremsstrahlung electrons in the \textit{Photon Tagger}.
$K^+$ were identified in the Forward Spectrometer via their momentum and $\beta$ determination. 
The photon from the decay $\Sigma^0\rightarrow \Lambda\gamma$ (labelled $\gamma'$ herein) was required to be identified in the BGO Rugby Ball.  Photons were rejected as $\gamma'$ candidates if the invariant mass of combinations of two photons was within 30\,MeV/c$^2$ of the $\pi^0$ mass (corresponding to $2\sigma$ of the mass resolution). The missing mass to the $K^+\gamma'$ system was determined for remaining $\gamma'$ candidates and the photon where this was the closest to the $\Lambda$ mass was identified. The four momentum of the remaining $\gamma'$ candidate was subsequently boosted into the rest frame of the $\Sigma^0$. 
Figure~\ref{fig:decayphoton} shows the $\gamma'$ energy in this frame, where the peak at 74\,MeV is consistent with the expected two-body decay momentum. 
The simulated $K^+\Sigma^0$ data exhibit a small, almost flat distribution under the peak arising from misidentified $\pi^0$ decay photons and neutrons from the subsequent $\Lambda$ decay.  
An additional combined background where forward $e^+$ and $\pi^+$ are misidentified as $K^+$ is also included.  The $e^+$ originated from pair production in the beam in random coincidence with  hardware triggers.  The $\pi^+$ contribution was relatively small and arose from other hadronic reactions at high momentum (over approximately 800\,MeV/c).  These distributions were generated by an equivalent analysis identifying forward going negatively charged particles, where the  $e^-$ and $\pi^-$ distributions are the same.  Two datasets were obtained, applying either a one or two sigma selection over the peak (69 to 79\,MeV or 62 to 86\,MeV respectively and approximating the peak as a Gaussian distribution).

The remaining candidate events were rejected if charged and neutral particle multiplicities exceeded either decay modes: $\Sigma^0\rightarrow \gamma\Lambda\rightarrow \gamma(\pi^0 n)$ or $\gamma(\pi^- p)$.
No requirement was made upon the detection of the $\Lambda$ decay particles as to do so significantly lowered the overall detection efficiency. 

\begin{figure} [h]
	\centering
	\includegraphics[width=\columnwidth,trim={0.0cm 0.0cm 0.0cm 0.0cm},clip=true]{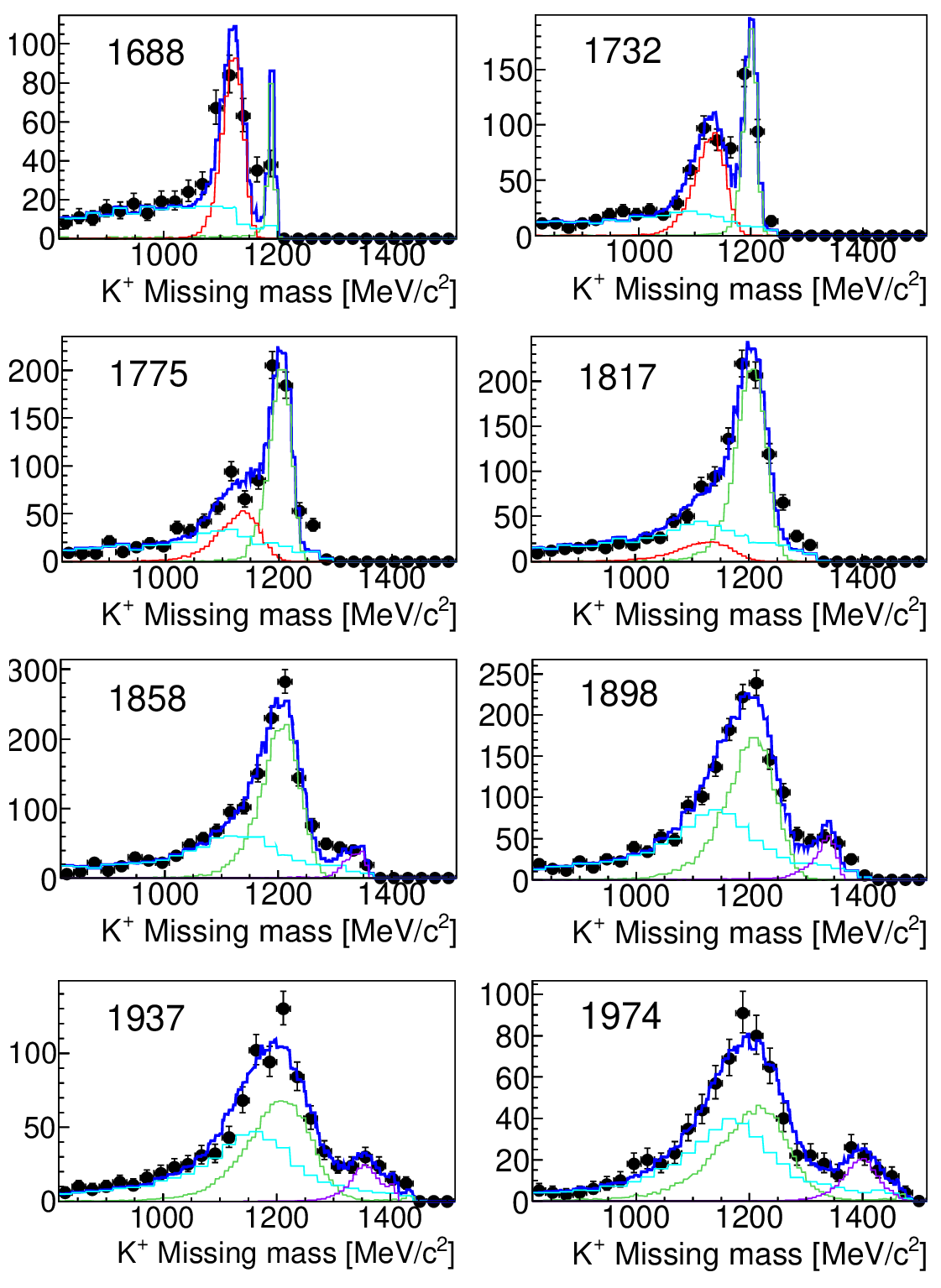}
	\caption{Missing mass recoiling from forward $K^+$ candidates after a  2$\sigma$ or 1$\sigma$ $\gamma'$  identification (below and above $W = 1930$\,MeV respectively).  Every third $E_\gamma$ interval is shown and labelled in MeV.  The Y-axes are counts per 24\,MeV/c$^2$ interval.  The spectra are fitted with simulated $K^+\Sigma^0$ (green line),  $K^+\Lambda$ (red line), $K^+\Sigma^0(1385)$ (purple line), and $e^+/\pi^+$ background (cyan line). The blue line is the summed total fit.}
	\label{fig:ksmissingmass}
\end{figure}

The missing mass recoiling from forward
$K^+$ after the $\gamma'$ identification is shown in Fig.~\ref{fig:ksmissingmass}. A fit was made using spectra from simulated $K^+\Lambda$, $K^+\Sigma^0$ and $K^+\Sigma^0(1385)$ events, and real data from misidentified $e^+/\pi^+$ described above \footnote{Simulated $K^+\Lambda(1405)$ spectra were not used due to the mass degeneracy to the $\Sigma^0(1385)$.}. 
The simulated data used distributions from previously measured cross sections, however the energy and angle intervals were sufficiently small so that the spectra depended solely on the resolution and response of the experimental setup. 
Simulated data of non open-strange final states including $\pi^+$ were also investigated.  The most dominant channel was $\pi^0\Delta^+$, however even this proved to be a negligible contribution and made no significant impact\footnote{Background from  $\pi^+\Delta^0$ had an even smaller contribution as the $\pi^+$ from this channel had a lower momentum and contributed less under the $K^+$ signal.}. The fit used separate scaling factors for each of the missing mass templates in order to extract the $K^+\Sigma^0$ yield.

\begin{figure} [h]
	\centering
	\resizebox{\columnwidth}{!}{%
		\includegraphics[trim={0 0 2cm 4.3cm},clip]{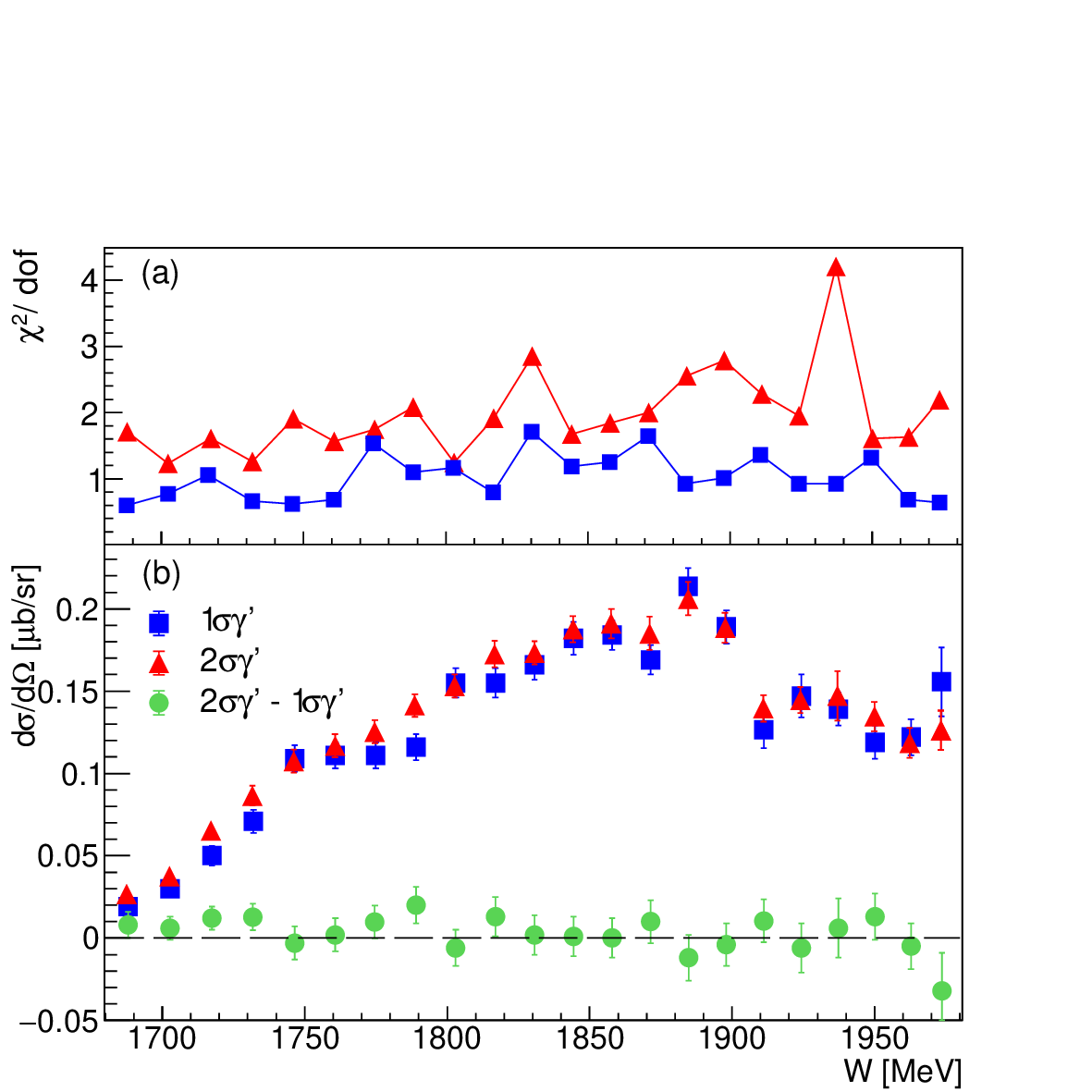}
	}
	\caption{(a) Reduced $\chi^2$ for each fit to the missing mass spectra shown in fig.~\ref{fig:ksmissingmass}.  The red triangles and blue squares correspond to the $1\sigma\gamma'$ and $2\sigma\gamma'$ selection respectively.  
	(b) The differential cross section for \kaonangle{}$> 0.9$ calculated using either the 1$\sigma\gamma'$ or 2$\sigma\gamma'$ selection.  The green circles are the difference.}
	\label{fig:comparecs}
\end{figure}

The $1\sigma\gamma'$ selection gave an expected lower yield of events by a factor of 2/3, however the signal to background improved by a factor of 3/2 for energies over $W = 1740$\,MeV.
  For the $2\sigma\gamma'$ selection above $W = 1930$\,MeV, the integral of the background exceeded the signal within the signal region.  To avoid additional systematic uncertainties which may have arisen from separating such large background contributions, the results presented in sec.~3 use the $2\sigma\gamma'$ data below $W = 1930$\,MeV, and the $1\sigma\gamma'$ data above.
The reduced $\chi^2$ of the fits to the $1\sigma\gamma'$ missing mass spectra are shown in fig.~\ref{fig:comparecs}(a).  The $1\sigma\gamma'$ missing mass spectra have an average reduced $\chi^2$ of 1.01 over the full energy range, and the fits to the $2\sigma\gamma'$ missing mass spectra have an average of 1.90 for $W$ below 1930\,MeV.
The cross sections calculated using both the $2\sigma\gamma'$ and $1\sigma\gamma'$ data are shown in fig.~\ref{fig:comparecs}(b).  There is good agreement over the full energy range, where the difference (the green filled circles) is consistent with zero.  

\begin{figure} [h]
	\centering
	\vspace*{0cm}
	\resizebox{\columnwidth}{!}{%
		\includegraphics{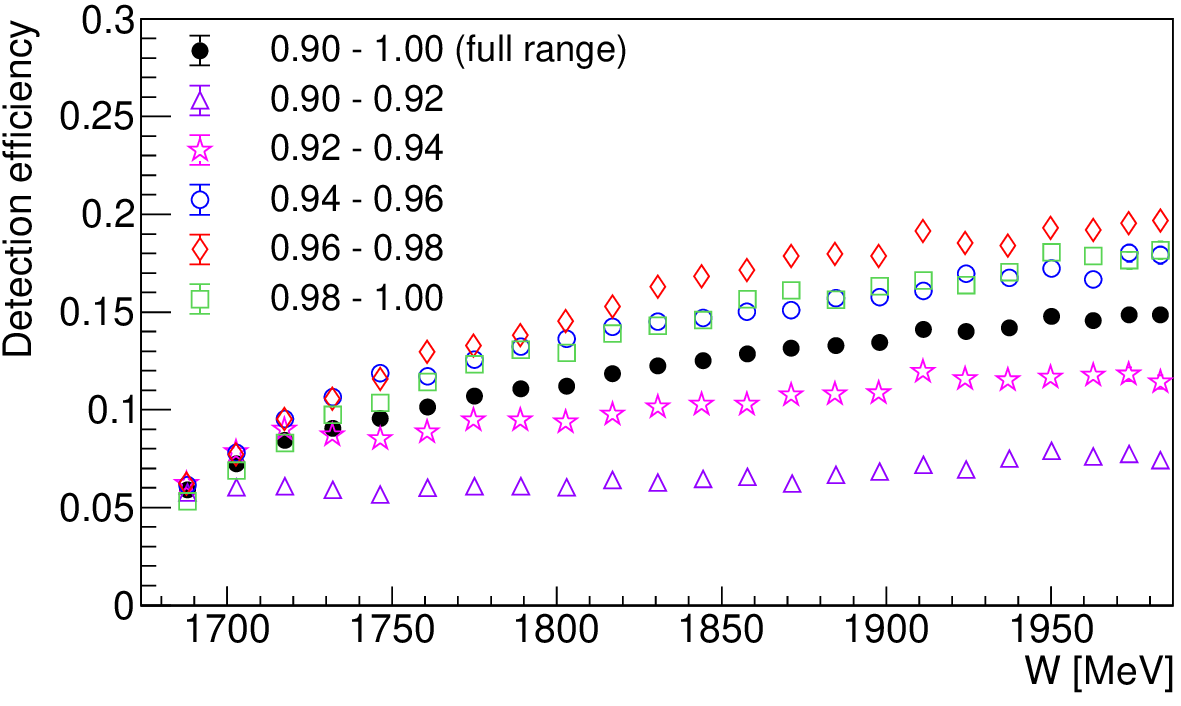}
	}
	\caption{Detection efficiency versus $W$ for the $2\sigma\gamma'$ selection.  The statistical error is typically 1\,\% for the full $0.90 <$\kaonangle{}$ < 1.00$ interval (thick black line with filled circles) and 2-3\,\% for the smaller \kaonangle{} intervals (open data points).
	}
	\label{fig:deteff}
\end{figure}

The detection efficiency shown in Fig.~\ref{fig:deteff}  includes the loss of approximately 50\% $K^+$ decaying mid-flight.  A smooth distribution is observed, with an increase in efficiency at the more forward \kaonangle{} intervals.  The detection efficiency was determined using a simulated model of the BGOOD experiment with Geant4~\cite{geant4}.  The simulation includes the full experimental geometry, including the Photon Tagger system, target holding structure, cladding material surrounding the crystals in the BGO Rugby Ball, and individual fibres and wires for the Forward Spectrometer detectors.  The Open Dipole magnetic field vector was both measured and simulated throughout the volume of the Open Dipole, including the fringe field beyond the magnet's yoke.  Energy and time resolutions for all detectors were determined and included, and the efficiency of MOMO, SciFi, the Drift Chambers and the TOF walls were measured as a function of particle $\beta$ and position from real data and applied to the simulation (see Ref.~\cite{technicalpaper} for details).  Including the individual detector efficiencies changes the net detection efficiency of forward going particles by the order of approximately 5\,\%. 

The efficiencies of the hardware triggers were also measured and included (see Ref.~\cite{klambdapaper} for details).  The trigger used for these results required approximately 80\,MeV energy deposition in the BGO Rugby Ball, and a forward track identified by timing coincidences between the SciFi and ToF detectors.  Due to the large time range and slight misalignment of the trigger windows, the efficiency of this trigger varied smoothly from 90\,\% to  95\,\% for forward particle $\beta$ of 0.58 and 0.92, corresponding to $W = 1688$ and 1973\,MeV respectively.

Cross sections of well known reactions have been previously measured to ensure an accurate understanding of the Forward Spectrometer geometry, the track finding algorithm efficiency and trigger efficiencies.  Ref.~\cite{technicalpaper} shows the differential cross section, for example, for $\gamma p \rightarrow \eta p$ when the proton is identified in the forward spectrometer. 
The measured cross section gave a good agreement to previous datasets.  The differential cross section was also shown versus the meson polar angle for four energy intervals over the N(1535)1/2$^-$ resonance.  The distribution was smooth and almost flat as would be expected, agreeing with PWA solutions.  The measured $\gamma p \rightarrow K^+\Lambda$ differential cross section using the same experimental setup is shown in Ref.~\cite{klambdapaper} and is of particular relevance as the $K^+$ detection was very similar to the analysis presented here.  The data exhibited a smooth, almost flat distribution over the Forward Spectrometer acceptance of $0.9 < $\kaonangle{} $< 1.0$, with good agreement to previous data at a slightly more backward \kaonangle{}.  No cusp-like structure was observed at the same $K^+$ momentum of 1\,GeV/c, as is presented in the results in section~3.  

Systematic uncertainties are the same as described in ref.~\cite{klambdapaper} and are divided into two components. The \textit{scaling uncertainty} is a constant fraction of the measured cross section, where the entire dataset would be expected to scale by a fixed value.  The dominant sources of this are the photon beam spot position and the photon flux normalisation, both estimated as 4\,\%.  The uncertainty in the $\gamma'$ identification of 1.5\,\% was determined by comparing the differential cross section for both the $1\sigma\gamma'$ and $2\sigma\gamma'$ selection over a small range of the photon beam energy where the signal to background ratio is high (approximately $W = 1770$ to 1870\,MeV). 
The \textit{fitting uncertainty} arises from extracting the number of events from the missing mass spectra.  As each fit to an energy and angle interval is independent of the others, this uncertainty permits the individual scaling to larger or smaller values for each data point. This was determined by additionally including simulated $\Delta^0\pi^+$ events in the background distribution (which proved to have a negligible effect) and by varying the fit range.  The fitting uncertainty is small over most energies, rising from 4\,\% to 8\,\% between $W =$ 1870 and 1920\,MeV.  From $W =$ 1920\,MeV to 1975\,MeV, the fitting uncertainty rises quickly to 28\,\% due to the missing mass distribution of the signal becoming broader and the yield reducing compared to the background. 
Data beyond $W = 1975$\,MeV were considered insufficiently accurate and rejected.

\section{Results and interpretation}

\begin{figure} [h]
	\centering
	\vspace*{0cm}
	\resizebox{\columnwidth}{!}{%
		\includegraphics{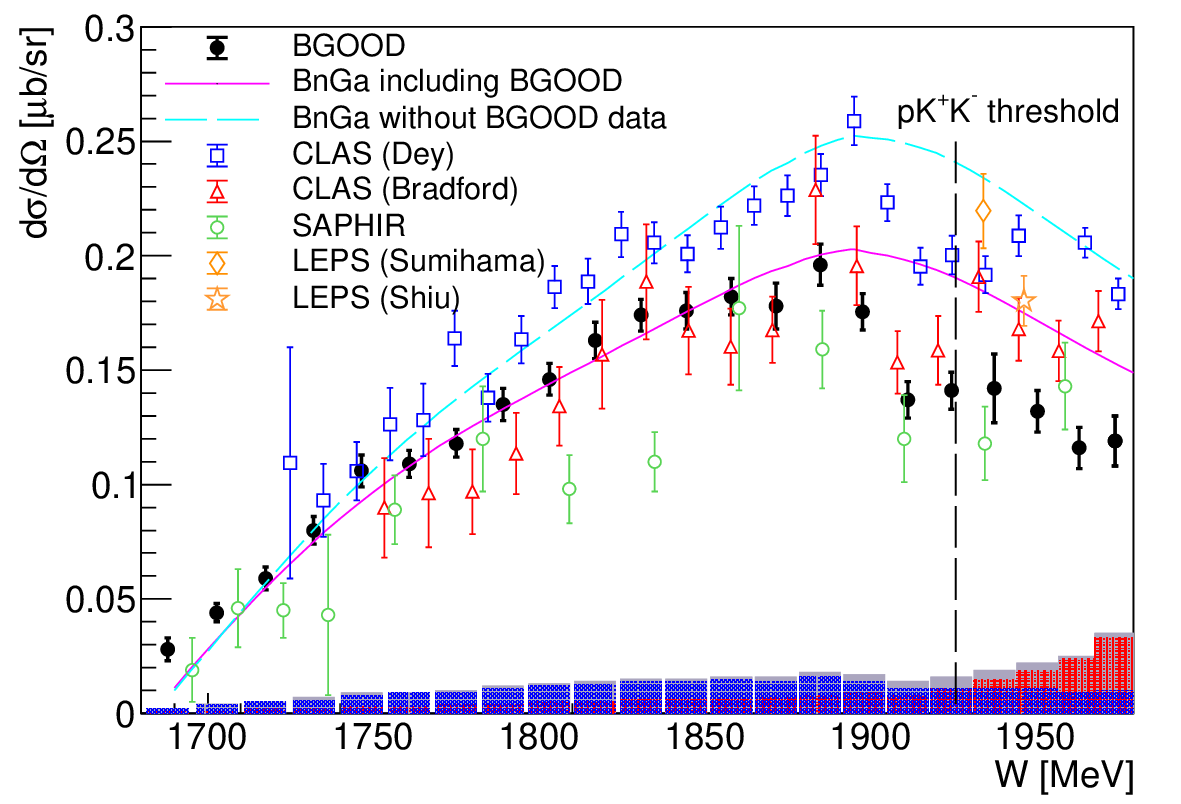}
	}
	\caption{$\gamma p \rightarrow K^+\Sigma^0$ differential cross section for \kaonangle $>0.90$	(black circles). The systematic uncertainties on the abscissa are in three components: The shaded blue, red and grey bars are the scaling, fitting and summed uncertainties respectively.  Data from Dey \textit{et al.} (CLAS)~\cite{dey10}, Bradford \textit{et al.} (CLAS)~\cite{bradford06}, Glander \textit{et al.} (SAPHIR)~\cite{glander04}, Sumihama \textit{et al.} (LEPS)~\cite{sumihama06} and Shiu \textit{et al.} (LEPS)~\cite{shiu18} are additionally shown and labelled in the legend.
	The CLAS data are at the more backward angle of $0.85 < \cos\theta_{CM}^{K} < 0.95$.  
	The Bonn-Gatchina PWA solutions~\cite{anisovich14} with and without the inclusion of the new data are the solid magenta and dashed cyan lines respectively.  The $pK^+K^-$ threshold is indicated by the dashed black line.
	}
	\label{fig:kscstotal}
\end{figure}

The $K^+\Sigma^0$ differential cross section for \kaonangle{} $> 0.9$ is shown in Fig.~\ref{fig:kscstotal}.  The CLAS datasets also shown are at the more backward angle of 0.85 $<$ \kaonangle{} $<$ 0.95, and the SAPHIR data are the only other from threshold at this most forward \kaonangle{} interval.  The $pK^-K^+$ threshold (which is close to the other thresholds mentioned in the introduction) is also shown.

This new dataset has the highest statistics from threshold to a centre-of-mass energy, $W = 1970$\,MeV.
The statistical error is approximately half of that of the SAPHIR~\cite{glander04} and CLAS data of Bradford \textit{et al.}~\cite{bradford06} and smaller than the CLAS data of Dey \textit{et al.}~\cite{dey10} below 1850\,MeV, above which it is comparable.

\begin{figure} [h]
	\centering
	\includegraphics[width=\columnwidth,trim={0cm 2.2cm 0cm 1.1cm},clip=true]{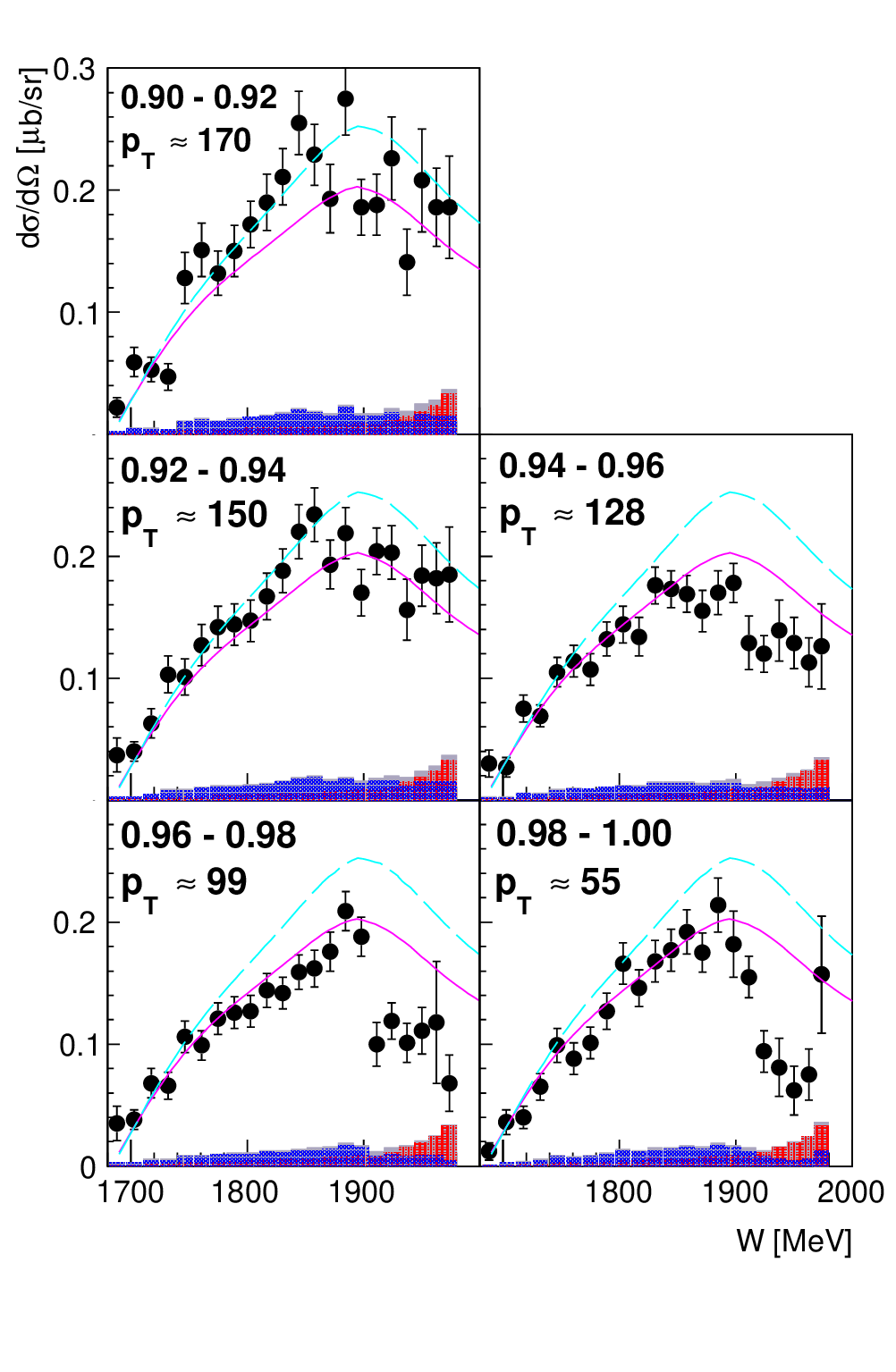}\\
	\caption{$\gamma p \rightarrow K^+\Sigma^0$ differential cross section for intervals of 0.02 in \kaonangle{} (labelled inset).
		The approximate $K^+$ transverse momentum at $W = 1900$\,MeV, $p_T$, is given for each interval in units of MeV/c (see Sec.~3 for details).
		The labelling of data points and fits is the same as in Fig.~\ref{fig:kscstotal}.
	}
	\label{fig:kscsvsenergyfine}
\end{figure}

\begin{figure} [h]
	\includegraphics[width=\columnwidth]{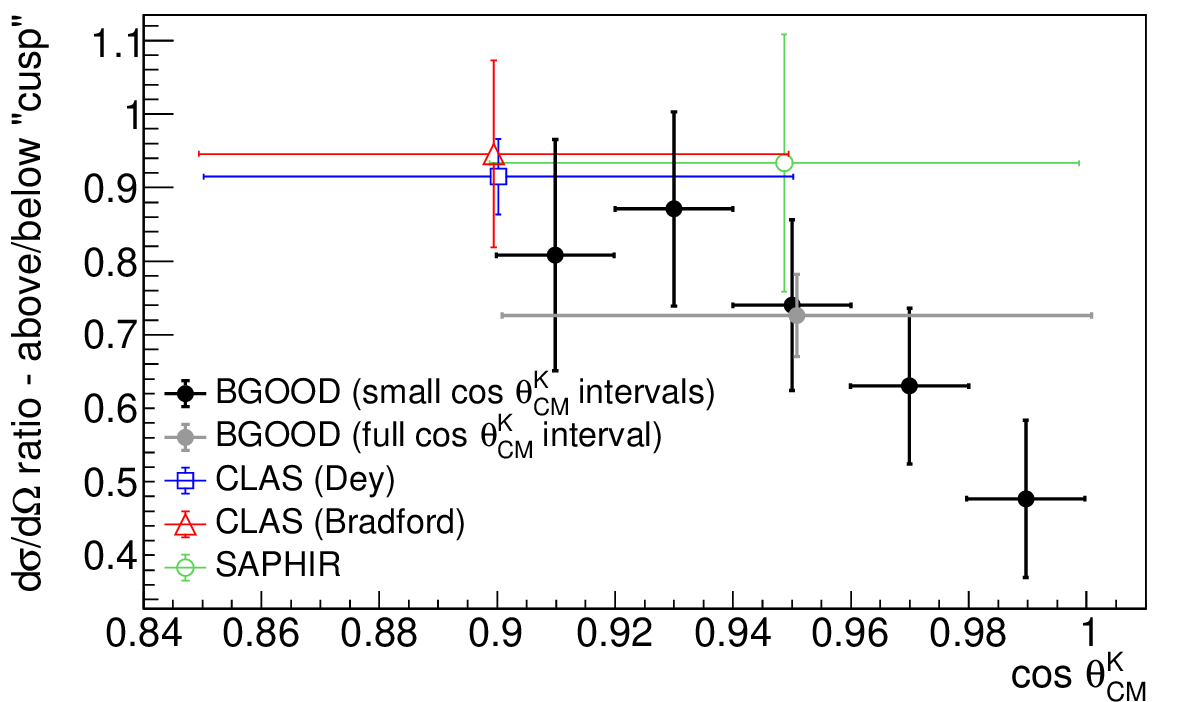}
	\caption{The ratio of the differential cross section from $W = 1924 $ to 1974\,MeV compared to $W = 1831$ to 1885\,MeV (above and below the cusp-like structure).  The data are the average of the differential cross section over these intervals, weighted by the statistical and systematic error.  The vertical error bars are the statistical uncertainties, the horizontal error bars are the interval in \kaonangle{} for the given dataset.    
	}
	\label{fig:kscsratio2}
\end{figure}

The drop at  $W = 1900$\,MeV  was regarded as a peak or a bump in the cross section in previous measurements.  Numerous PWA and isobar model solutions attributed this to a
N(1895)3/2$^-$, N(1900)1/2$^+$, N(1910)1/2$^+$ and N(1900)3/2$^-$~\cite{mart99,janssen02,corthals06,anisovich07} resonance for example, but with no firm agreement.
The improved statistics and \kaonangle{} forward acceptance of this new dataset however allow a cusp-like structure to be resolved at this energy, where the cross section drops by approximately one-third over a 20\,MeV range.
The high \kaonangle{} resolution for BGOOD at forward angles permits the fine binning shown in Fig.~\ref{fig:kscsvsenergyfine}, where the differential cross section is in 0.02 \kaonangle{} intervals versus $W$.  The cusp becomes more pronounced at the most forward angle interval, \kaonangle$> 0.98$, where there is a reduction of over 50\,\% over a 30\,MeV range.  At the most backward intervals,  the cusp is difficult to discern, starting to only become visible around $0.94 <$\kaonangle$< 0.96$.
This is evident in Fig.~\ref{fig:kscsratio2} where the ratio of the average cross section immediately before and after the cusp is shown.  The same ratio is shown for the CLAS and SAPHIR datasets.  The extent of the cusp changes quickly with \kaonangle{}, demonstrating that the CLAS datasets and this new data are compatible and that the cusp only becomes visible at the acceptance limit of CLAS.

Given the quickly changing strength of the cusp and the different \kaonangle{} intervals for the datasets, only a comparison between the data below  the cusp at $W = 1900$\,MeV can be sensibly made.  Within this $W$ region,  there is an agreement between these data and the Bradford CLAS data to within approximately 5\,\%, which is smaller than the combined systematic uncertainty of 13\,\%.  
The Dey CLAS data are higher by approximately 20\,\%, which is larger than the combined systematic uncertainty of 15\,\% and the SAPHIR data are 25\,\% lower, which is nearly double the combined systematic uncertainty of 14\,\%. 

The Bonn-Gatchina BG2019 solution~\cite{anisovich14} (the dashed cyan line in Figs.~\ref{fig:kscstotal} and \ref{fig:kscsvsenergyfine}) was constrained by a combined fit to both CLAS datasets.\footnote{This was achieved by weighting the data according to the statistical error.  The algorithm generally allows a change of weighting factors by a factor up to 1.4 to obtain optimal combined fits to previous and newly introduced data.}  The fit obtained a reduced $\chi^2$ of 6.03.  
After including the BGOOD data and refitting all couplings to resonant contributions and $t$ and $u$ channel exchange amplitudes and including a refit of reactions in the Bonn-Gatchina database, a notable improvement of the reduced $\chi^2$ to 2.08 was achieved (the solid magenta line), with 
no significant changes to the fit occurring at  more backward angles (\kaonangle{}$< 0.85$) covered by the CLAS data.  It should be stressed that the fit is still dominated by the CLAS results given the larger amount of data points available compared to BGOOD.  
Additional resonant contributions were  iteratively included to test for further improvements.  A $\Delta(1917)5/2^-$ with a relatively narrow width of 59\,MeV gave the best improvement to these data (not shown in the figures), only influencing the most forward data points.  This improvement however cannot adequately describe the cusp at 1900\,MeV.

\begin{figure} [h]
	\centering
	\vspace*{0cm}
	\resizebox{0.9\columnwidth}{!}{%
		\includegraphics{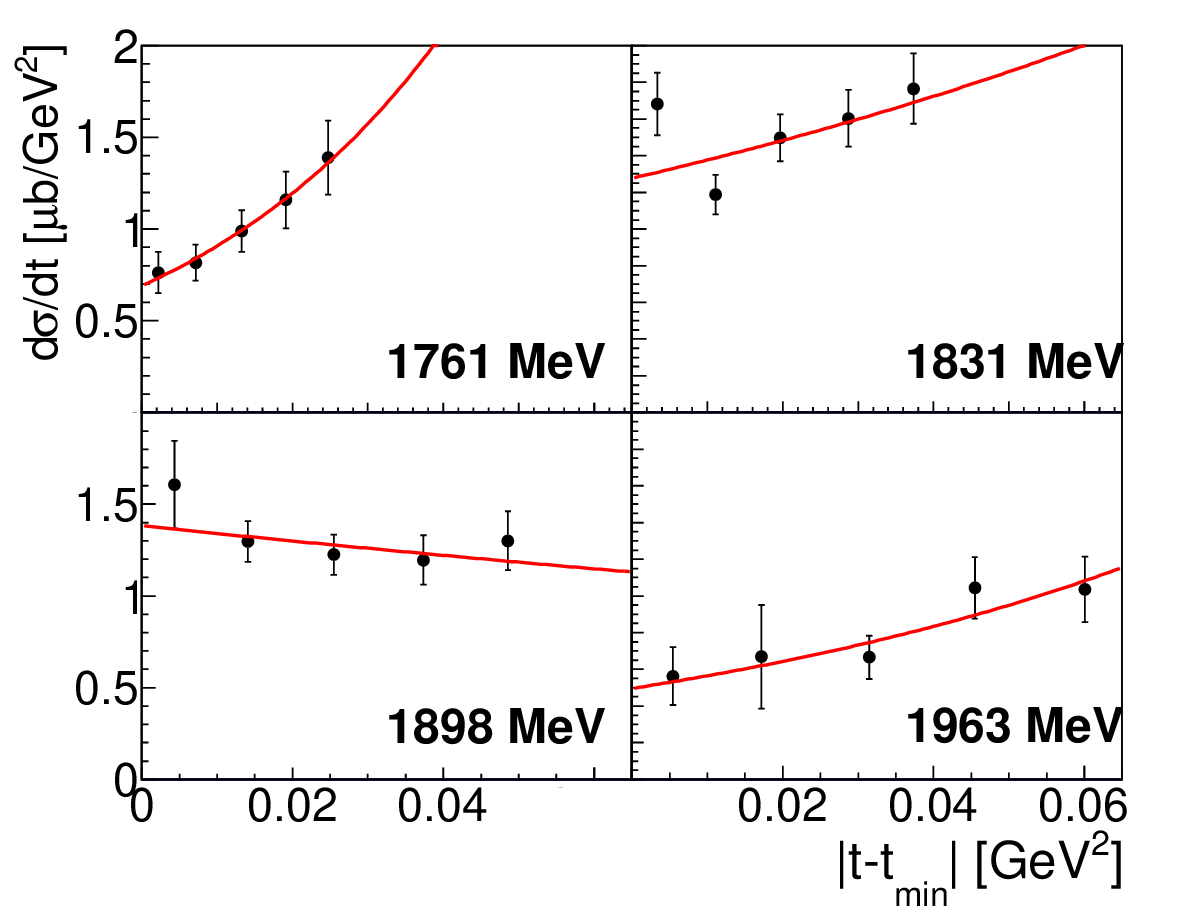}
		.	}
	\caption{ $\gamma p \rightarrow K^+\Sigma^0$ d$\sigma$/d$t$ versus $|t - t_\mathrm{min}|$ for intervals of $W$, labelled inset.  Only the statistical error is shown and included in the fit.  The red lines are eq.~1 fitted to the data.
	}
	\label{fig:ksfittingslope}
\end{figure}

The strong dependence on \kaonangle{} and momentum exchange may suggest threshold dynamics or meson-baryon interactions play a  significant role in the reaction mechanism.
  If an experimental observation was due to extended loosely bound structures, for example meson-baryon type multi-quark configurations, a strong dependence upon momentum transfer in $t$-channel production processes could be expected. 
To investigate such behaviour, the data were determined as a function of the Mandelstam variable, $t$.  
To extrapolate the data to the minimum possible momentum transfer, $t_\mathrm{min}$, where the $K^+$ has a polar angle of $0^\circ$, $t$ was determined for each $W$ and \kaonangle{} interval, examples of which are shown in Fig.~\ref{fig:ksfittingslope} and fitted with the function in eq.~1.  Only the statistical error of each data point was included in the fit.  

\begin{equation}
\frac{\mathrm{d}\sigma}{\mathrm{d}t} = \frac{\mathrm{d}\sigma}{\mathrm{d}t}\Big|_{t=t_\mathrm{min}}e^{S|t-t_\mathrm{min}|}
\end{equation}\label{eq:fitfunction}


\begin{figure} [h]
	\centering
	\includegraphics[width=\columnwidth,trim={0.0cm 0.0cm  0cm 0.0cm},clip=true]{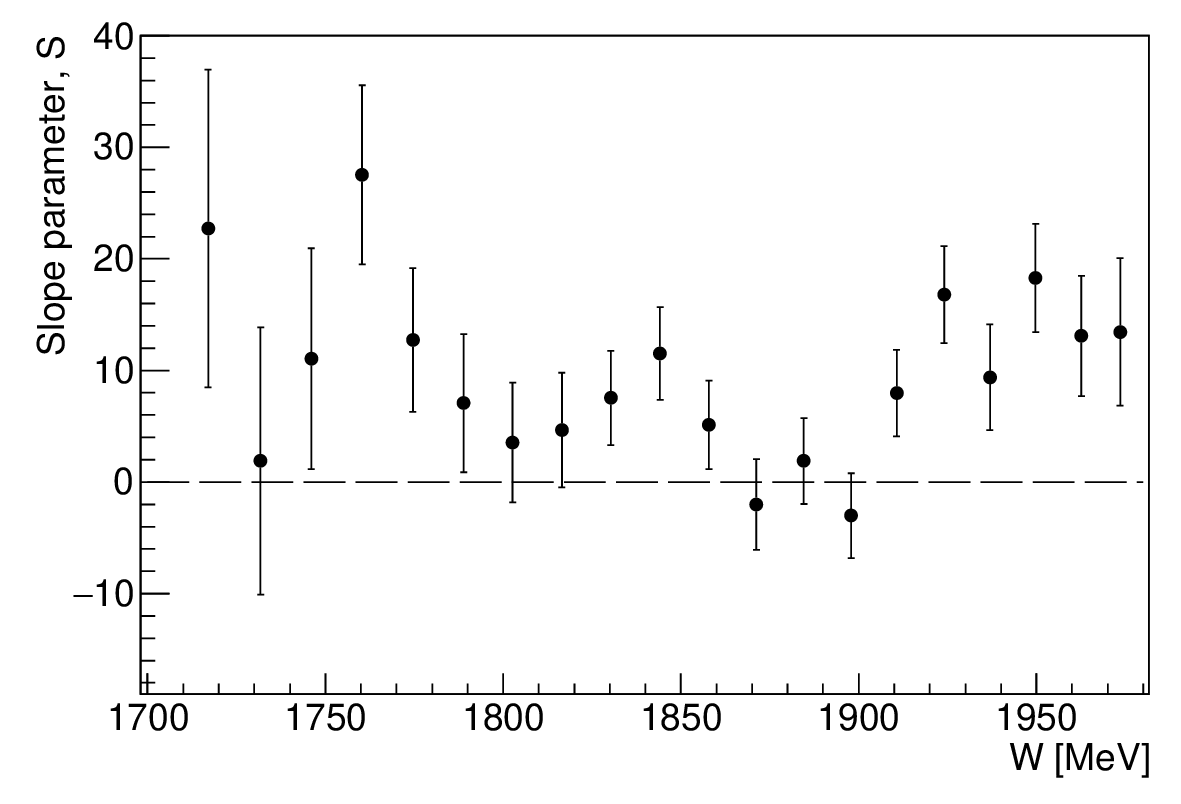}
	\caption{The slope parameter, $S$ (in eq.~1) versus $W$.
	}
	\label{fig:slope}
\end{figure}

The fits to the $t$ dependence for each $W$ interval were used to determine $d\sigma/dt|_{t=t_\mathrm{min}}$ and $S$. 
The slope parameter, $S$, shown in Fig.~\ref{fig:slope} is positive and appears flat (although with limited statistical precision) for approximately the first 150\,MeV from threshold, indicative of $s$-channel contributions.  The fact that this occurs over a larger $W$ range compared to $K^+\Lambda$ shown in Ref.~\cite{klambdapaper} may be due to both $N^*$ and $\Delta^*$ resonance contributions.
$S$ begins to drop approximately 50\,MeV before the cusp and becomes zero or slightly negative at 1900\,MeV, as would be expected for a dominating $t$-channel process. 
Above 1900\,MeV, there is a sharp rise back to positive values of $S$,  where it then remains flat.  This quick change with respect to $W$ may indicate that a significant $t$-channel contribution is lost and could be interpreted as a threshold effect, where an off-shell contribution becomes on-shell above $W = 1900$\,MeV.

\begin{figure} [h]
	\centering	
	\includegraphics[width=\columnwidth,trim={0.0cm 0.0cm  0cm 0.0cm},clip=true]{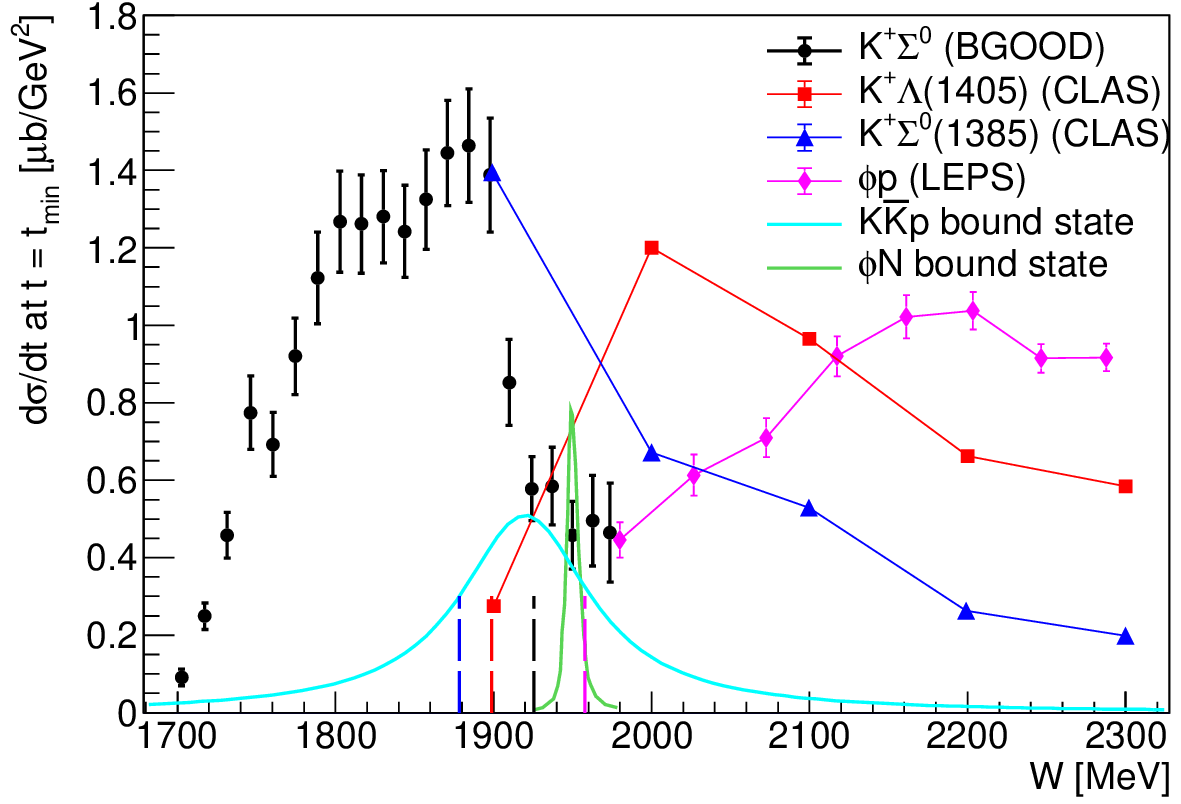}
	
	\caption{$\gamma p \rightarrow K^+\Sigma^0$ differential cross section, d$\sigma$/d$t$ extrapolated to $t_\mathrm{min}$ versus $W$ (filled black circles).
		Data from the CLAS Collaboration for the $K^{+}\Lambda(1405)$ and $K^{+}\Sigma^0(1385)$ final states are included, which were estimated from ref.~\cite{moriyathesis} (red squares and blue triangles respectively).  The magenta diamonds are the $\phi p$ cross section from the LEPS collaboration~\cite{mibe05}.  The vertical dashed lines indicate the respective thresholds, with the addition of the $K^+K^-p$ threshold indicated by the black dashed line.  The predictions of the $K\bar{K}N$~\cite{torres09} and $\phi N$~\cite{guo17} bound states are shown as the cyan and green lines at arbitrary scales.
	}
	\label{fig:kscstmin}
\end{figure}

The differential cross section extrapolated to $t_\mathrm{min}$ in Fig.~\ref{fig:kscstmin} exhibits a particularly pronounced drop in strength at $W = 1900$\,MeV.
The following subsections discuss these data in relation to the predictions and evidence of bound hadronic states outlined in the introduction.  

\subsection{The $X$(2000) pentaquark candidate}
 This new dataset is in a similar kinematic regime to the diffractive $K^+\Sigma^0$ production reported by the Sphinx Collaboration~\cite{golovkin95}.  The proposed $X(2000)$ became only apparent in the Sphinx Collaboration analysis when it was required that the $K^+$ (or $\Sigma^0$) transverse momentum, $p_T$ was smaller than 141\,MeV/c.
 For the BGOOD data, $p_T$ at the cusp is labelled in Fig.~\ref{fig:kscsvsenergyfine} for each \kaonangle{} interval.  $p_T$ is approximately three times smaller in the most forward angle interval than what could be accessed with the CLAS data, where it becomes comparable to the Fermi momentum of the deuteron.
The quickly changing strength of the cusp with respect to $p_T$ could suggest an extended, loosely bound meson-baryon system where the constituents travel parallel at very low $p_T$.  The proposed width of 91\,MeV/c$^2$ ensures that the $X$(2000) would significantly overlap the observed cusp.

\subsection{Predicted hadronic bound states}
Superimposed on Fig.~\ref{fig:kscstmin} are the predicted $K\bar{K}p$ and $\phi p$ bound states~\cite{torres09,guo17}.  The  $K\bar{K}p$ state mass of 1920\,MeV is exactly at the cusp, and the width of 100\,MeV (which is an approximation) appears comparable to the width of the cusp.

The predicted bound $N\phi$ state~\cite{guo17} has a mass of approximately 1950\,MeV, which appears immediately after the cusp.  Ref.~\cite{guo17} proposed the state could be produced experimentally using a gold target, where due to Fermi momentum the $\phi$ could have sufficiently low relative momentum to another spectator nucleon to form a bound system.  At the cusp at $W=1900$, the three momentum component of $t$ is very similar  (between 500 to 550\,MeV/c) to the  momentum at the maximum amplitude of the bound state formation (see fig.~4 in ref.~\cite{guo17}).

The predicted strange pentaquark consisting of a bound $\Sigma(1385)K$ state is calculated to have a mass of 1873\, and a binding energy of 7.4\,MeV~\cite{huang18}, lying approximately 20\,MeV below the cusp.

\subsection{Near threshold behaviour of channels with open and hidden strangeness}

Cross section data for $K^+\Lambda(1405)$, $K^+\Sigma^0(1385)$ and $\phi p$ at $t_\mathrm{min}$ are also shown on fig.~\ref{fig:kscstmin}.  The $K^+Y$ channels are from CLAS~\cite{moriyathesis} (an approximation from extrapolating graphical data), and the $p\phi$ data are from the LEPS collaboration~\cite{mibe05}. 
As all of the predicted bound states described in the previous section have $I = \frac{1}{2}$, the  $K^+\Sigma^0(1385)$ has been multiplied by the corresponding Clebsch-Gordan coefficient.

 All four channels have thresholds close to the cusp.  There appears to be an almost ``smooth" transition between the $K^+\Sigma^0$ and $p\phi$ cross sections.   There also seems to be a similar behaviour of the $K^+\Sigma^0(1385)$ channel compared to  $K^+\Sigma^0$, where a pronounced drop is observed at a similar $W$.  Qualitatively, it could be argued that a conserved quantity of $s\bar{s}$ constituent quark pairs is distributed across the different channels and  bound states.

\section{Conclusions}

Differential cross sections for  $\gamma p \rightarrow K^+\Sigma^0$ for \kaonangle $> 0.9$ have been measured from threshold to $W = 1970$\,MeV. 
A strong asset of this new dataset is the forward angular acceptance up to \kaonangle = 1.0, exceeding previous CLAS measurements and having only been previously achieved with SAPHIR at a lower precision.
With the large dataset and fine resolution in \kaonangle{}, the BGOOD experiment is able to resolve a cusp-like structure at $W = 1900$\,MeV.
No firm conclusions can yet be made, however the behaviour may be indicative of re-scattering effects close to open and hidden strange thresholds in an energy region where there are multiple predictions of hadronic bound states.

\section*{Acknowledgements}

We would like to thank the staff and shift-students of the ELSA accelerator for providing an excellent beam.  We thank Eulogio Oset, \`{A}ngels Ramos  and Reinhard Schumacher for insightful comments and Eberhard Klempt for help with the Bonn-Gatchina PWA. 

This work is supported by SFB/TR-16, DFG project numbers 388979758 and 405882627, the RSF grant number 19-42-04132, and the Third Scientific Committee of the INFN.  This publication is part of a project that has received funding from the European Union’s Horizon 2020 research and innovation programme under grant agreement STRONG – 2020 - No.~824093.
P.~L.~Cole gratefully acknowledges the support from the U.S. National Science Foundation (NSF-PHY-1307340, NSF-PHY-1615146, and NSF-PHY-2012826) and the Fulbright U.S. Scholar Program (2014/2015).
\bibliographystyle{unsrt}

\newpage
\appendix
\section{Tabulated data}
\begin{table}[h]
			\begin{tabular}{ c c c c c c c}
			\hline\hline
			\multicolumn{7}{l}{$0.90<$\kaonangle$<1.00$} \\
			\hline
			$W$ &  $\Delta W$ & d$\sigma$/d$\Omega$ & $\delta_\mathrm{stat}$ &$\delta_\mathrm{sys}$&  $\delta_\mathrm{scaling}$ &  $\delta_\mathrm{fitting}$ \\ 	 
			
			MeV & MeV &$\mu$b/sr & $\mu$b/sr & $\mu$b/sr & $\mu$b/sr & $\mu$b/sr\\
			\hline
1688.0	&	14.9	&	0.028	&	0.005	&	0.002	&	0.002	&	0.001	\\
1702.8	&	14.7	&	0.044	&	0.004	&	0.004	&	0.004	&	0.001	\\
1717.4	&	14.6	&	0.059	&	0.005	&	0.005	&	0.005	&	0.002	\\
1732.0	&	14.5	&	0.080	&	0.006	&	0.007	&	0.006	&	0.002	\\
1746.4	&	14.4	&	0.106	&	0.007	&	0.009	&	0.008	&	0.003	\\
1760.8	&	14.3	&	0.109	&	0.006	&	0.009	&	0.009	&	0.003	\\
1775.0	&	14.1	&	0.118	&	0.006	&	0.010	&	0.009	&	0.004	\\
1789.0	&	14.1	&	0.135	&	0.007	&	0.012	&	0.011	&	0.004	\\
1803.0	&	13.9	&	0.146	&	0.007	&	0.013	&	0.012	&	0.005	\\
1816.9	&	13.8	&	0.163	&	0.008	&	0.014	&	0.013	&	0.005	\\
1830.7	&	13.7	&	0.174	&	0.007	&	0.015	&	0.014	&	0.006	\\
1844.3	&	13.6	&	0.176	&	0.008	&	0.015	&	0.014	&	0.006	\\
1857.9	&	13.5	&	0.182	&	0.008	&	0.016	&	0.015	&	0.006	\\
1871.4	&	13.5	&	0.178	&	0.010	&	0.016	&	0.014	&	0.007	\\
1884.7	&	13.3	&	0.196	&	0.009	&	0.018	&	0.016	&	0.008	\\
1898.0	&	13.2	&	0.175	&	0.008	&	0.017	&	0.014	&	0.009	\\
1911.2	&	13.1	&	0.137	&	0.008	&	0.014	&	0.011	&	0.008	\\
1924.3	&	13.1	&	0.141	&	0.008	&	0.016	&	0.011	&	0.011	\\
1937.3	&	12.9	&	0.142	&	0.015	&	0.019	&	0.011	&	0.015	\\
1950.2	&	12.9	&	0.132	&	0.009	&	0.022	&	0.011	&	0.019	\\
1962.8	&	12.3	&	0.116	&	0.009	&	0.025	&	0.009	&	0.024	\\
1973.7	&	9.5	&	0.119	&	0.011	&	0.035	&	0.010	&	0.033	\\
		\hline\hline
\end{tabular}\caption{$\gamma p \rightarrow K^+\Sigma^0$ differential cross section data (d$\sigma$/d$\Omega$) for $0.90<$\kaonangle$<1.00$.  The median and width of each centre-of-mass interval are labelled $W$ and $\Delta W$respectively.  The statistical, systematic, and the two components of the systematic uncertainties (scaling and fitting) are labelled $\delta_\mathrm{stat}$, $\delta_\mathrm{sys}$, $\delta_\mathrm{scaling}$ and $\delta_\mathrm{fitting}$ respectively.}\label{table:cs1}
\end{table}


\begin{table}[h]
	\begin{tabular}{ c c c c c c c}
		\hline\hline
		\multicolumn{7}{l}{$0.90<$\kaonangle$<0.92$} \\
		\hline
		$W$ &  $\Delta W$ & d$\sigma$/d$\Omega$ & $\delta_\mathrm{stat}$ &$\delta_\mathrm{sys}$&  $\delta_\mathrm{scaling}$ &  $\delta_\mathrm{fitting}$ \\ 	 
		
		MeV & MeV &$\mu$b/sr & $\mu$b/sr & $\mu$b/sr & $\mu$b/sr & $\mu$b/sr\\
		\hline
1688.0	&	14.9	&	0.022	&	0.008	&	0.002	&	0.002	&	0.001	\\
1702.8	&	14.7	&	0.059	&	0.012	&	0.005	&	0.005	&	0.001	\\
1717.4	&	14.6	&	0.053	&	0.010	&	0.005	&	0.004	&	0.002	\\
1732.0	&	14.5	&	0.047	&	0.011	&	0.004	&	0.004	&	0.002	\\
1746.4	&	14.4	&	0.128	&	0.021	&	0.011	&	0.010	&	0.003	\\
1760.8	&	14.3	&	0.151	&	0.022	&	0.013	&	0.012	&	0.003	\\
1775.0	&	14.1	&	0.132	&	0.018	&	0.011	&	0.011	&	0.004	\\
1789.0	&	14.1	&	0.150	&	0.021	&	0.013	&	0.012	&	0.004	\\
1803.0	&	13.9	&	0.172	&	0.019	&	0.014	&	0.014	&	0.005	\\
1816.9	&	13.8	&	0.190	&	0.023	&	0.016	&	0.015	&	0.005	\\
1830.7	&	13.7	&	0.211	&	0.023	&	0.018	&	0.017	&	0.006	\\
1844.3	&	13.6	&	0.255	&	0.026	&	0.021	&	0.020	&	0.006	\\
1857.9	&	13.5	&	0.229	&	0.025	&	0.019	&	0.018	&	0.006	\\
1871.4	&	13.5	&	0.193	&	0.028	&	0.017	&	0.015	&	0.007	\\
1884.7	&	13.3	&	0.275	&	0.030	&	0.024	&	0.022	&	0.008	\\
1898.0	&	13.2	&	0.186	&	0.023	&	0.017	&	0.015	&	0.009	\\
1911.2	&	13.1	&	0.188	&	0.025	&	0.017	&	0.015	&	0.008	\\
1924.3	&	13.1	&	0.226	&	0.034	&	0.021	&	0.018	&	0.011	\\
1937.3	&	12.9	&	0.141	&	0.027	&	0.019	&	0.011	&	0.015	\\
1950.2	&	12.9	&	0.208	&	0.042	&	0.025	&	0.017	&	0.019	\\
1962.8	&	12.3	&	0.186	&	0.032	&	0.028	&	0.015	&	0.024	\\
1973.7	&	9.5	&	0.186	&	0.042	&	0.037	&	0.015	&	0.033	\\
		\hline\hline
\end{tabular}\caption{$\gamma p \rightarrow K^+\Sigma^0$ differential cross section data for $0.90<$\kaonangle$<0.92$.  The notation is the same as in table~\ref{table:cs1}.}

	\begin{tabular}{ c c c c c c c}
			&&&&&&\\
		\hline\hline
		\multicolumn{7}{l}{$0.92<$\kaonangle$<0.94$} \\
		\hline
		$W$ &  $\Delta W$ & d$\sigma$/d$\Omega$ & $\delta_\mathrm{stat}$ &$\delta_\mathrm{sys}$&  $\delta_\mathrm{scaling}$ &  $\delta_\mathrm{fitting}$ \\ 	 
		
		MeV & MeV &$\mu$b/sr & $\mu$b/sr & $\mu$b/sr & $\mu$b/sr & $\mu$b/sr\\
		\hline
1688.0	&	14.9	&	0.037	&	0.014	&	0.003	&	0.003	&	0.001	\\
1702.8	&	14.7	&	0.040	&	0.008	&	0.003	&	0.003	&	0.001	\\
1717.4	&	14.6	&	0.063	&	0.012	&	0.005	&	0.005	&	0.002	\\
1732.0	&	14.5	&	0.103	&	0.015	&	0.009	&	0.008	&	0.002	\\
1746.4	&	14.4	&	0.101	&	0.015	&	0.009	&	0.008	&	0.003	\\
1760.8	&	14.3	&	0.127	&	0.017	&	0.011	&	0.010	&	0.003	\\
1775.0	&	14.1	&	0.142	&	0.017	&	0.012	&	0.011	&	0.004	\\
1789.0	&	14.1	&	0.144	&	0.017	&	0.012	&	0.011	&	0.004	\\
1803.0	&	13.9	&	0.147	&	0.017	&	0.013	&	0.012	&	0.005	\\
1816.9	&	13.8	&	0.167	&	0.019	&	0.014	&	0.013	&	0.005	\\
1830.7	&	13.7	&	0.188	&	0.018	&	0.016	&	0.015	&	0.006	\\
1844.3	&	13.6	&	0.220	&	0.022	&	0.019	&	0.018	&	0.006	\\
1857.9	&	13.5	&	0.234	&	0.022	&	0.020	&	0.019	&	0.006	\\
1871.4	&	13.5	&	0.193	&	0.020	&	0.017	&	0.015	&	0.007	\\
1884.7	&	13.3	&	0.219	&	0.021	&	0.019	&	0.018	&	0.008	\\
1898.0	&	13.2	&	0.170	&	0.019	&	0.016	&	0.014	&	0.009	\\
1911.2	&	13.1	&	0.204	&	0.019	&	0.018	&	0.016	&	0.008	\\
1924.3	&	13.1	&	0.203	&	0.022	&	0.020	&	0.016	&	0.011	\\
1937.3	&	12.9	&	0.156	&	0.025	&	0.019	&	0.012	&	0.015	\\
1950.2	&	12.9	&	0.184	&	0.025	&	0.024	&	0.015	&	0.019	\\
1962.8	&	12.3	&	0.182	&	0.029	&	0.028	&	0.015	&	0.024	\\
1973.7	&	9.5	&	0.185	&	0.039	&	0.037	&	0.015	&	0.033	\\
	\hline\hline
\end{tabular}\caption{$\gamma p \rightarrow K^+\Sigma^0$ differential cross section data for $0.92<$\kaonangle$<0.94$.  The notation is the same as in table~\ref{table:cs1}.}
\end{table}
		\begin{table}[h]
			\begin{tabular}{ c c c c c c c}
				\hline\hline
				\multicolumn{7}{l}{$0.94<$\kaonangle$<0.96$} \\
				\hline
				$W$ &  $\Delta W$ & d$\sigma$/d$\Omega$ & $\delta_\mathrm{stat}$ &$\delta_\mathrm{sys}$&  $\delta_\mathrm{scaling}$ &  $\delta_\mathrm{fitting}$ \\ 	 
				
				MeV & MeV &$\mu$b/sr & $\mu$b/sr & $\mu$b/sr & $\mu$b/sr & $\mu$b/sr\\
				\hline
1688.0	&	14.9	&	0.030	&	0.011	&	0.003	&	0.002	&	0.001	\\
1702.8	&	14.7	&	0.027	&	0.008	&	0.003	&	0.002	&	0.001	\\
1717.4	&	14.6	&	0.075	&	0.011	&	0.006	&	0.006	&	0.002	\\
1732.0	&	14.5	&	0.069	&	0.009	&	0.006	&	0.005	&	0.002	\\
1746.4	&	14.4	&	0.105	&	0.012	&	0.009	&	0.008	&	0.003	\\
1760.8	&	14.3	&	0.114	&	0.013	&	0.010	&	0.009	&	0.003	\\
1775.0	&	14.1	&	0.107	&	0.013	&	0.009	&	0.009	&	0.004	\\
1789.0	&	14.1	&	0.132	&	0.014	&	0.011	&	0.011	&	0.004	\\
1803.0	&	13.9	&	0.144	&	0.015	&	0.012	&	0.012	&	0.005	\\
1816.9	&	13.8	&	0.134	&	0.016	&	0.012	&	0.011	&	0.005	\\
1830.7	&	13.7	&	0.176	&	0.015	&	0.015	&	0.014	&	0.006	\\
1844.3	&	13.6	&	0.173	&	0.015	&	0.015	&	0.014	&	0.006	\\
1857.9	&	13.5	&	0.169	&	0.015	&	0.015	&	0.014	&	0.006	\\
1871.4	&	13.5	&	0.155	&	0.017	&	0.014	&	0.012	&	0.007	\\
1884.7	&	13.3	&	0.170	&	0.018	&	0.016	&	0.014	&	0.008	\\
1898.0	&	13.2	&	0.178	&	0.016	&	0.017	&	0.014	&	0.009	\\
1911.2	&	13.1	&	0.129	&	0.022	&	0.013	&	0.010	&	0.008	\\
1924.3	&	13.1	&	0.120	&	0.015	&	0.015	&	0.010	&	0.011	\\
1937.3	&	12.9	&	0.139	&	0.025	&	0.018	&	0.011	&	0.015	\\
1950.2	&	12.9	&	0.129	&	0.021	&	0.022	&	0.010	&	0.019	\\
1962.8	&	12.3	&	0.113	&	0.020	&	0.025	&	0.009	&	0.024	\\
1973.7	&	9.5	&	0.126	&	0.035	&	0.035	&	0.010	&	0.033	\\
		\hline\hline
	\end{tabular}\caption{$\gamma p \rightarrow K^+\Sigma^0$ differential cross section data for $0.94<$\kaonangle$<0.96$.  The notation is the same as in table~\ref{table:cs1}.}
	\begin{tabular}{ c c c c c c c}
		&&&&&&\\
		\hline\hline
		\multicolumn{7}{l}{$0.96<$\kaonangle$<0.98$} \\
		\hline
		$W$ &  $\Delta W$ & d$\sigma$/d$\Omega$ & $\delta_\mathrm{stat}$ &$\delta_\mathrm{sys}$&  $\delta_\mathrm{scaling}$ &  $\delta_\mathrm{fitting}$ \\ 	 
		
		MeV & MeV &$\mu$b/sr & $\mu$b/sr & $\mu$b/sr & $\mu$b/sr & $\mu$b/sr\\
		\hline
1688.0	&	14.9	&	0.035	&	0.014	&	0.003	&	0.003	&	0.001	\\
1702.8	&	14.7	&	0.038	&	0.008	&	0.003	&	0.003	&	0.001	\\
1717.4	&	14.6	&	0.068	&	0.012	&	0.006	&	0.005	&	0.002	\\
1732.0	&	14.5	&	0.066	&	0.011	&	0.006	&	0.005	&	0.002	\\
1746.4	&	14.4	&	0.106	&	0.013	&	0.009	&	0.008	&	0.003	\\
1760.8	&	14.3	&	0.099	&	0.012	&	0.009	&	0.008	&	0.003	\\
1775.0	&	14.1	&	0.121	&	0.013	&	0.010	&	0.010	&	0.004	\\
1789.0	&	14.1	&	0.126	&	0.013	&	0.011	&	0.010	&	0.004	\\
1803.0	&	13.9	&	0.127	&	0.013	&	0.011	&	0.010	&	0.005	\\
1816.9	&	13.8	&	0.144	&	0.014	&	0.013	&	0.012	&	0.005	\\
1830.7	&	13.7	&	0.142	&	0.013	&	0.013	&	0.011	&	0.006	\\
1844.3	&	13.6	&	0.159	&	0.014	&	0.014	&	0.013	&	0.006	\\
1857.9	&	13.5	&	0.162	&	0.015	&	0.015	&	0.013	&	0.006	\\
1871.4	&	13.5	&	0.176	&	0.016	&	0.016	&	0.014	&	0.007	\\
1884.7	&	13.3	&	0.209	&	0.016	&	0.019	&	0.017	&	0.008	\\
1898.0	&	13.2	&	0.188	&	0.016	&	0.017	&	0.015	&	0.009	\\
1911.2	&	13.1	&	0.100	&	0.018	&	0.012	&	0.008	&	0.008	\\
1924.3	&	13.1	&	0.119	&	0.015	&	0.015	&	0.010	&	0.011	\\
1937.3	&	12.9	&	0.101	&	0.016	&	0.017	&	0.008	&	0.015	\\
1950.2	&	12.9	&	0.111	&	0.019	&	0.021	&	0.009	&	0.019	\\
1962.8	&	12.3	&	0.118	&	0.050	&	0.025	&	0.009	&	0.024	\\
1973.7	&	9.5      &	 0.068	 &	 0.023	&	0.034	&	0.005	&	0.033	\\
	\hline\hline
\end{tabular}\caption{$\gamma p \rightarrow K^+\Sigma^0$ differential cross section data for $0.96<$\kaonangle$<0.98$.  The notation is the same as in table~\ref{table:cs1}.}
\end{table}

\begin{table}[h]
	\begin{tabular}{ c c c c c c c}
		\hline\hline
		\multicolumn{3}{l}{$0.98<$\kaonangle$<1.00$} \\
		\hline
		$W$ &  $\Delta W$ & d$\sigma$/d$\Omega$ & $\delta_\mathrm{stat}$ &$\delta_\mathrm{sys}$&  $\delta_\mathrm{scaling}$ &  $\delta_\mathrm{fitting}$ \\ 	 
		
		MeV & MeV &$\mu$b/sr & $\mu$b/sr & $\mu$b/sr & $\mu$b/sr & $\mu$b/sr\\
	
		\hline
1688.0	&	14.9	&	0.012	&	0.007	&	0.001	&	0.001	&	0.001	\\
1702.8	&	14.7	&	0.036	&	0.010	&	0.003	&	0.003	&	0.001	\\
1717.4	&	14.6	&	0.040	&	0.009	&	0.004	&	0.003	&	0.002	\\
1732.0	&	14.5	&	0.065	&	0.011	&	0.006	&	0.005	&	0.002	\\
1746.4	&	14.4	&	0.099	&	0.014	&	0.009	&	0.008	&	0.003	\\
1760.8	&	14.3	&	0.088	&	0.013	&	0.008	&	0.007	&	0.003	\\
1775.0	&	14.1	&	0.101	&	0.013	&	0.009	&	0.008	&	0.004	\\
1789.0	&	14.1	&	0.127	&	0.015	&	0.011	&	0.010	&	0.004	\\
1803.0	&	13.9	&	0.166	&	0.017	&	0.014	&	0.013	&	0.005	\\
1816.9	&	13.8	&	0.146	&	0.015	&	0.013	&	0.012	&	0.005	\\
1830.7	&	13.7	&	0.168	&	0.017	&	0.015	&	0.013	&	0.006	\\
1844.3	&	13.6	&	0.177	&	0.017	&	0.015	&	0.014	&	0.006	\\
1857.9	&	13.5	&	0.192	&	0.018	&	0.017	&	0.015	&	0.006	\\
1871.4	&	13.5	&	0.175	&	0.016	&	0.016	&	0.014	&	0.007	\\
1884.7	&	13.3	&	0.214	&	0.022	&	0.019	&	0.017	&	0.008	\\
1898.0	&	13.2	&	0.182	&	0.027	&	0.017	&	0.015	&	0.009	\\
1911.2	&	13.1	&	0.155	&	0.017	&	0.015	&	0.012	&	0.008	\\
1924.3	&	13.1	&	0.094	&	0.017	&	0.013	&	0.008	&	0.011	\\
1937.3	&	12.9	&	0.081	&	0.024	&	0.016	&	0.006	&	0.015	\\
1950.2	&	12.9	&	0.062	&	0.020	&	0.020	&	0.005	&	0.019	\\
1962.8	&	12.3	&	0.075	&	0.021	&	0.024	&	0.006	&	0.024	\\
1973.7	&	9.5	     &	 0.157	 &	 0.048	 &	 0.036	 &	 0.013	 &	 0.033	\\
		\hline\hline
	\end{tabular}\caption{$\gamma p \rightarrow K^+\Sigma^0$ differential cross section data for $0.98<$\kaonangle$<1.00$.  The notation is the same as in table~\ref{table:cs1}.}
\end{table}

\begin{table}[h]
	\centering
	\begin{tabular}{ c c S[table-format=3.2] S[table-format=3.2] c  c}
		\hline\hline
		$W$ &  $\Delta W$ & $S$ & $\delta S_\mathrm{stat}$  &  d$\sigma/$d$t$   &   $\delta$d$\sigma/$d$t_\mathrm{stat}$ \\
		
		MeV & MeV &   &  &  $\mu$b/GeV$^2$   &   $\mu$b/GeV$^2$ \\
		\hline
1702.8	&	14.7	&	75.86	&	37.94	&	0.092	&	0.022	\\
1717.4	&	14.6	&	22.72	&	14.25	&	0.253	&	0.035	\\
1732.0	&	14.5	&	1.89	&	11.98	&	0.459	&	0.059	\\
1746.4	&	14.4	&	11.05	&	9.90	&	0.778	&	0.095	\\
1760.8	&	14.3	&	27.53	&	8.02	&	0.695	&	0.083	\\
1775.0	&	14.1	&	12.72	&	6.45	&	0.922	&	0.098	\\
1789.0	&	14.1	&	7.07	&	6.17	&	1.123	&	0.118	\\
1803.0	&	13.9	&	3.55	&	5.38	&	1.271	&	0.130	\\
1816.9	&	13.8	&	4.66	&	5.15	&	1.263	&	0.127	\\
1830.7	&	13.7	&	7.53	&	4.21	&	1.284	&	0.119	\\
1844.3	&	13.6	&	11.52	&	4.14	&	1.245	&	0.119	\\
1857.9	&	13.5	&	5.13	&	3.97	&	1.328	&	0.129	\\
1871.4	&	13.5	&	-2.00	&	4.07	&	1.446	&	0.136	\\
1884.7	&	13.3	&	1.89	&	3.84	&	1.468	&	0.147	\\
1898.0	&	13.2	&	-3.02	&	3.78	&	1.389	&	0.148	\\
1911.2	&	13.1	&	8.00	&	3.87	&	0.856	&	0.111	\\
1924.3	&	13.1	&	16.79	&	4.33	&	0.581	&	0.083	\\
1937.3	&	12.9	&	9.39	&	4.74	&	0.585	&	0.100	\\
1950.2	&	12.9	&	18.27	&	4.86	&	0.459	&	0.087	\\
1962.8	&	12.3	&	13.09	&	5.38	&	0.498	&	0.117	\\
1973.7	&	9.5	&	13.46	&	6.61	&	0.467	&	0.128	\\
		\hline\hline
	
	\end{tabular}\caption{The slope parameter, $S$ and the differential cross section extrapolated to $t=t_\mathrm{min}$, d$\sigma/$d$t$ (eq.~1) shown in figs.~9 and 10 respectively.}
\end{table}

\end{document}